\documentclass[pre,twocolumn,showpacs,preprintnumbers]{revtex4}
\pdfoutput=1
\usepackage[plainpages=false, colorlinks=true, anchorcolor=blue, linkcolor=blue, citecolor=blue, bookmarks=false]{hyperref}
\usepackage{amsfonts,amsmath,amssymb}
\usepackage{graphicx}
\usepackage{color}
\usepackage{natbib}
\graphicspath{c:/Users/landau/Downloads}
\pdfoutput=1
\usepackage[dvipsnames]{xcolor}

\begin{document}

\title{Do $\tau$ Lepton Branching fractions obey Benford's Law?}
\author{Aisha  \surname{Dantuluri}$^1$} \altaffiliation{E-mail: ep14btech11001@iith.ac.in}
\author{Shantanu   \surname{Desai}$^1$} \altaffiliation{E-mail: shantanud@iith.ac.in}

%
\affiliation{$^{1}$Department of Physics, IIT Hyderabad, Kandi, Telangana-502285, India}
\date{Received: date}
%
\begin{abstract}
According to Benford's law, the most significant digit in many datasets is not uniformly distributed, but obeys a well defined power law distribution with smaller digits appearing more often.  Among one of the myriad  particle physics datasets available, we find that the leading decimal digit for the $\tau$ lepton branching fraction \textcolor{black}{shows marginal disagreement with the  logarithmic behavior expected from the Benford distribution.} We quantify the deviation from Benford's law using a $\chi^2$ function valid for binomial data,  and obtain a $\chi^2$ value of 16.9  for nine degrees of freedom, which gives a  $p$-value of \textcolor{black}{about 5\%}, corresponding to a 1.6$\sigma$ disagreement. We also checked that the disagreement persists under scaling the branching fractions, as well as by redoing the analysis in a numerical system with a base different from 10. \textcolor{black}{Among all the digits, `9' shows the largest
discrepancy with an excess of $4\sigma$. This discrepancy is because the digit `9' is repeated  for three distinct groups of correlated modes, with each group having a frequency of two or three, leading to double-counting.  If we count each group of correlated modes only once, the discrepancy for this digit also disappears and we get pristine agreement with Benford distribution.}
\end{abstract}

%

\maketitle
\section{Introduction}
\label{sec:intro}
Many naturally occurring distributions tend to adhere to a logarithmic distribution as predicted by Benford's law~\cite{Pinkham,Hill98}. Though it may appear to be coincidence, there is more depth to the law than initially apparent.  The law states that given a distribution of numbers, the fraction of numbers with leading digit $d$ is given by~\cite{newcomb,benford}:

\[P(n) = \log(1+\frac{1}{d})\]

The notion first arose in 1881, when Simon Newcomb \cite{newcomb}, an astronomer, realized that the first few pages of his book of logarithmic tables, containing numbers that started with `1', were far more worn and frayed than the later pages, suggesting that numbers with `1' as a leading digit occurred more frequently than the others. The phenomenon was again brought to the forefront by the physicist Frank Benford 57 years later, and Benford tested the hypothesis that Newcomb had suggested in his initial paper on datasets from a diverse variety domains, each of varying sizes, and found that in many cases, the data complied with the logarithmic rule.

Benford \cite{benford} himself, though credited for the formulation of the law, was of the opinion that it was an absurd event that was the result of `anomalous' and `outlaw' numbers. He tested a variety of datasets, ranging from numbers in addresses and newspapers, to physical constants. His studies resulted in him stating  that it would not hold for conventional sets of numbers, such as atomic weights and specific heats of compounds and materials, as his distributions showed, but it was a good fit for quantities such as the numerals from the front pages of newspapers. 

Over the following years, the distribution has garnered a considerable amount of interest from a variety of specialists, from an assortment of fields, such as  biology~\cite{bio1,bio2}, finance~\cite{finance1,finance2}, geophysics~\cite{geo},  seismology~\cite{seis}, spectroscopy~\cite{whyman,bormashenko} and astrophysics~\cite{Moret,astro2,astro3}. This law also has been used for practical applications in detecting banking frauds~\cite{finance2}. A great number of instances where the law holds have been discovered, often accompanied by attempts to explain these seemingly disconnected occurrences. 

Newcomb's initial interpretation and subsequent explanation, remains the simplest of the many such explanations that are present \cite{newcomb}. He argued that all numbers in a distribution can be expressed in the form \(10^{n+f}\) where \(n\) is an integer and \(f\)  is a fraction. Clearly, from the above definition, the value of the first significant digit is  dependent only on the first digit of the number \(10^f\). Newcomb mathematically intuits that  the fraction $f$ should be distributed uniformly along the interval [0,1) and thus we arrive at what is known today as Benford's distribution for the first significant digit, with each digit occurring with the aforementioned probability.

Many years later, Raimi attempted to provide a more rigorous mathematical proof \cite{raimi}, as did Hill \cite{hill}, who also showed that the law was invariant of base and scale, meaning that the units of the calculated values did not affect the distribution \cite{Hillb}. However, no explanation has yet been able to fully account for the phenomenon in its entirety. 

In this work, we study whether the branching fractions of the tau lepton (hereafter $\tau$),  which is an elementary sub-atomic and leptonic particle about  3500 times heavier than the electron,  obey Benford's law. The outline of this paper is as follows. Section~\ref{sec:ppreview} reviews previous applications of this law
to ancillary nuclear and particle physics datasets. Our analysis and results are described in Section~\ref{sec:analysis}. Section~\ref{sec:corr} addresses some discrepancies and possible reasons for these. Our conclusions are presented in Section~\ref{sec:conclusions}.

\section{Benford's Law on Particle Physics Datasets}
\label{sec:ppreview}
It is to be expected that the large availability of experimental and calculated data from particle physics compiled by the Particle Data Group~\cite{pdg}  could provide an extensive and diverse assortment of quantities to potentially test for agreement with Benford's law.

Previously, different quantities pertaining to particle physics and nuclear physics have been shown to be in good agreement with Benford's distribution. These quantities include the calculated and experimental half lives of $\alpha$ decay, as shown by Buck et. al~\cite{buck} , the experimental values of $\beta$ decay half lives as shown by Ni et. al~\cite{ni}, and the values of full hadron widths as shown by Shao et. al~\cite{Shao}. This was also recently confirmed by Farkas et. al with more data~\cite{Farkas}.  These works establish the applicability of Benford's law to the general area of particle physics, suggesting that there maybe some reigning physical phenomenon governing these natural processes and  a connection between these results. However, we note that so far there is no first principles physics explanation of why they show remarkable agreement with Benford's law. Rather, it has been shown by Farkas et. al \cite{Farkas} that there is no such underlying physical significance between the results.

\begin{figure}
  \includegraphics[width = \linewidth,scale=0.5]{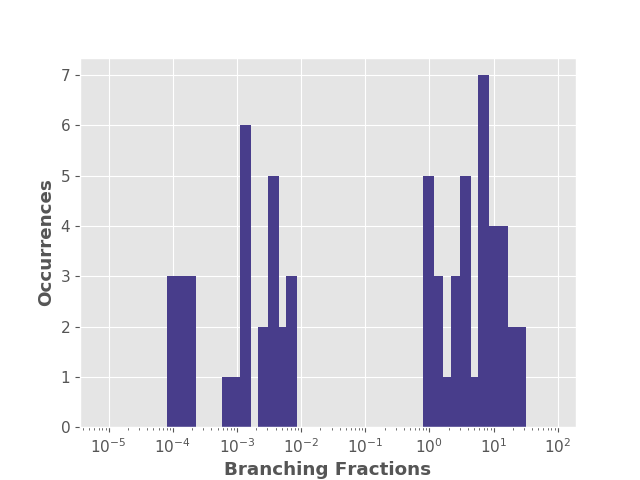}
  \caption{\textcolor{black}{ The above histogram indicates the distribution of the $\tau$ lepton branching fractions over a range of magnitudes. Initially 69 measurements are considered, avoiding those measurements which have large values of error in measurement.}}
  \label{fig:boat1}
\end{figure}

It is to be noted that all the quantities studied up until this point, to the best of our knowledge, are those in relation to the lifetime of particles including half lives and full widths of assorted particles. The range of the data examined has variations spanning \textcolor{black}{five} orders of magnitude. In the present situation we restrict our study to the branching fractions of the $\tau$ lepton. The data includes all reported values of branching fractions, regardless of the decay mechanism and process. The values of its branching fractions also range over as many as five orders of magnitude, from values as small as \(10^{-5}\) up to \(10^{1}\), \textcolor{black}{thus making it a suitable case for investigating for adherance to Benford's law. Fig.~\ref{fig:boat1} depicts the distribution of the first 69 compiled measurements on a log scale.} 

Moreover, like $\beta$ decay, $\tau$ branching is also governed by weak interactions. However, it is difficult to predict whether the values of $\tau$ lepton branching fractions will follow the Benford's distribution as the other analyzed quantities have. Therefore, in this work we test if the branching fractions of $\tau$ lepton obey Benford's law using 69 compiled measurements.

\section{Tau Branching Fractions and Benford's Law}
\label{sec:analysis}
A branching fraction is the fraction of particles which decay into a specific possible decay mode with respect to the total number of particles. There are many possible decay modes for the $\tau$ lepton. As the $\tau$ lepton is the heaviest amongst the leptons, it has the distinction of being the only lepton which can decay into hadrons as well. As mentioned before, the governing forces in $\tau$ decay are weak interactions.

\subsection{Data}
\label{ss:data}
The branching fractions for the $\tau$ lepton have been obtained from the Particle Data Group's Review and Summary tables \cite{pdg}. Each reported value is considered as a separate data point. The values provided in the tables include measurements with large error bars or upper  limits, corresponding to the  decay modes yet to be observed. Data points where only the upper limiting values of the branching fractions are provided, and data points with large errors, which would affect the significant digit,  are currently omitted from the analysis, to avoid uncertainty.

\begin{figure}
  \includegraphics[width = \linewidth,scale=0.5]{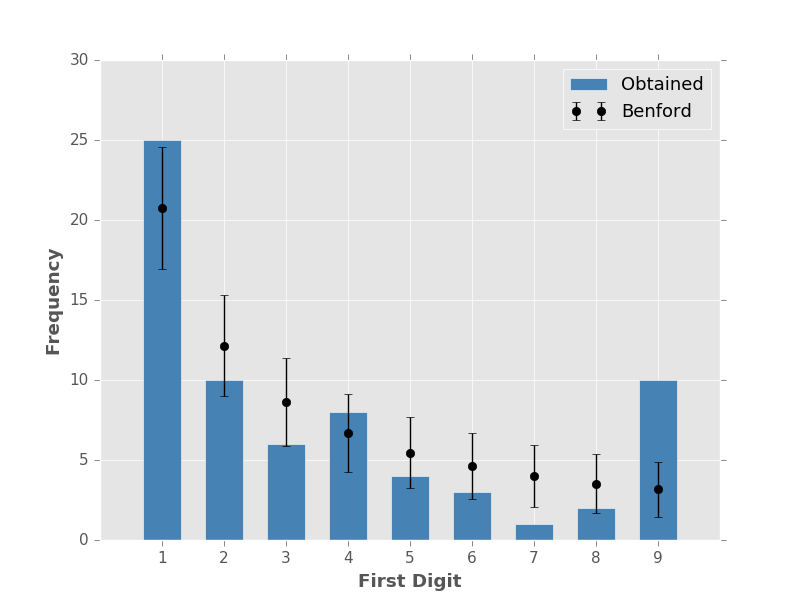}
  \caption{The bar graph indicates the frequency distribution of the first digit of the $\tau$ lepton branching fraction in terms of the fraction of total occurrences for the initially considered 69 measurements, avoiding those measurements which have large values of error in measurement. The values expected by Benford's law are graphed alongside, with the error bars indicating the binomial root mean square error.}
  \label{fig:boat2}
\end{figure}

With the modified data we simply count the number of data points for each possible first significant digit. In Table 1 we display the results obtained as opposed to the expected values. Fig.~\ref{fig:boat2} is a graphical representation of the same, where we plot a bar graph of the observed probability of occurrence of each significant digit, along with the predicted distribution. We also evaluate, tabulate and graph the expected root mean square error $[\Delta N(d)]$. As in references \cite{buck,ni} we use a binomial distribution to calculate the error, as either a digit $d$ occurs as the first significant digit, or it does not. Thus,
$$[\Delta N(d)] = \sqrt{NP(d)(1-P(d))}$$

Although qualitatively the distribution seems to agree well with the general trend expected from Benford's law, the observed probability overshoots for some digits and undershoots for others. We now quantitatively test these departures from a perfect Benford distribution in the next section.

\begin{table}

\begin{center}
\begin{tabular}[t]{|l ||c|| r|}
\hline\hline
 Digit & Count & Expected \\ [0.5ex]
 \hline
 1 & 25 & 20.77 $\pm$ 3.81\\
 2 & 10 & 12.15 $\pm$ 3.16\\
 3 & 6  & 8.62 $\pm$ 2.74\\
 4 & 8  & 6.68 $\pm$ 2.45\\
 5 & 4  & 5.46 $\pm$ 2.24\\
 6 & 3  & 4.61 $\pm$ 2.08\\
 7 & 1  & 4.00 $\pm$ 1.94\\
 8 & 2  & 3.52 $\pm$ 1.83\\
 9 & 10 & 3.15\ $\pm$ 1.73\\
\hline\hline
\end{tabular}
\end{center}
\caption{69 data points with negligible error bars are considered, and the distribution of the first significant digits is determined. We observe the difference of the obtained distribution in Column 2 to the expected distribution in Column 3. Column 3 also includes the expected binomial root mean square error expected for the distribution. We note that the distribution deviates from Benford's law, though the digits are not uniformly distributed either.}
\label{table:1}
\end{table}

\subsection{Evaluating confidence levels}
To quantify the deviation of the resulting distribution of the first significant digit for the branching ratios from the actual Benford's distribution, we evaluate Pearson's $\chi ^{2}$~\cite{Pearson} as a simple measure of goodness of fit. Previously, Shao et. al~\cite{Shao} have used this same statistic as an indicator of agreement of their data with Benford's distribution. On the other hand, Refs.~\cite{buck,ni} use a regular $\chi^2$ function~\cite{Press}. \textcolor{black}{However, later we shall see that for this case we cannot strictly apply Pearson or the regular $\chi^2$, because of too few data points in some of the bins. Nevertheless, in order to compare our results with previous particle physics related works on Benford's law, we first  quantify the disagreement using these metrics and then use a  $\chi^2$ function, which is tailor-made for our dataset.}

If $N_{B}(d)$ indicates the expected count of occurrences as according to Benford's law  and $N_{O}(d)$ is the  observed number of occurrences for a digit $d$, the Pearson $\chi ^{2}$ is defined as:
\begin{equation}
\chi ^{2} =\sum_{d=1}^{9} \frac{\left(N_{B}(d)-N_{O}(d)\right)^{2}} {N_{B}(d)} 
\end{equation}
On the other hand, in the regular $\chi^2$ statistic, the denominator is replaced by the root mean square error.
\begin{equation}
\chi ^{2} =\sum_{d=1}^{9} \left(\frac{N_{B}(d)-N_{O}(d)}{[\Delta N(d)]}\right)^{2} 
\end{equation}

If we evaluate the $\chi^{2}$ for the obtained distributions as given above, we find that we arrive at a Pearson $\chi^{2}$ value of 21.0, and a regular $\chi^{2}$ of 22.55. Statistically, the Pearson $\chi^{2}$ is said to have eight degrees of freedom, whereas the regular $\chi^2$ has nine degrees of freedom. For a perfect agreement with Benford's law, the ratio of $\chi^2$ to total degrees of freedom should be about unity, whereas we find a ratio of 2.62 for Pearson $\chi^{2}$ and 2.50 for regular $\chi^{2}$ .
For a distribution with eight degrees of freedom, the Pearson $\chi^{2}$ value obtained corresponds to a $p$-value of 0.007 using $\chi^2$ goodness of fit tables~\cite{Press}. For a distribution with nine degrees of freedom, the regular $\chi^{2}$ value also corresponds to about the same  $p$-value of 0.007. 

\textcolor{black}{Although the Pearson $\chi^2$ is widely used in literature for quantifying the goodness of fit for any model to binned data, one condition which must be fulfilled is that 80\% of the bins must have an expected value greater than five~\cite{Wall}. As we can see from Fig.~\ref{fig:boat2}, this is not satisfied. So strictly speaking in this case, we cannot use the $p$-value from the Pearson $\chi^2$ for quantifying the disagreement with Benford's law. Since the distribution within the bins follows the binomial distribution, we use the  $\chi^2$ valid for binomial distribution, which is defined as follows~\cite{Cousins}:}
\begin{equation}
\chi^2= 2 \sum \limits_{d=1}^9 N_{O}(d) \ln\left( \frac{N_O (d)}{N_B (d)}\right)
\label{eq:binomial}
\end{equation}

\textcolor{black}{To see where the $\chi^2$ in the above equation comes out, one first constructs a likelihood ratio given by the ratio of the likelihood of the observed data given the model to that of the maximum likelihood estimate given the data. It follows from Wilk's theorem~\cite{Wilks} that this likelihood ratio asymptotically follows the $\chi^2$ distribution. This $\chi^2$ can be used for both parameter estimation and goodness of fit testing. More details on the conceptual underpinnings behind Eq.~\ref{eq:binomial} can be found in Refs.~\cite{Cousins,Cash}. When we evaluate this $\chi^2$, we obtain a value of 16.9 for 9 degrees of freedom, which corresponds to a $p$-value of 0.05.}

\textcolor{black}{The $p$-value indicates the  probability that we would get a value larger than the observed $\chi^2$ distribution, when the null hypothesis (in this case, agreement with Benford's law) is true.} 
To convert this $p$-value to significance, we adopt the usual convention of  calculating the number of standard deviations a Gaussian variable would fluctuate on one side to give the same $p$-value~\cite{Ganguly}. \textcolor{black}{This $p$-value of 0.05 corresponds to 1.6$\sigma$ disagreement with Benford's law, which is very marginal.}

Therefore, in summary we find that the first significant digit of $\tau$ branching fractions  shows a very slight disagreement with  Benford's law, at a confidence level of approximately \textcolor{black}{95\%, corresponding to a 1.6$\sigma$ discrepancy, which is not very significant.}
\label{ss:ecl}

\subsection{Analyzing Ambiguous Data}
\label{ss:se}

\begin{figure}
  \includegraphics[width = \linewidth,scale=0.5]{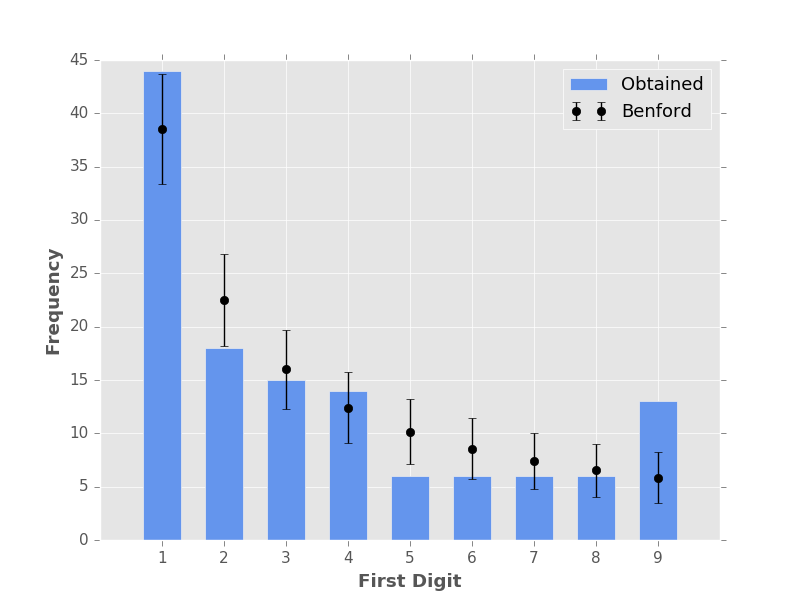}
  \caption{The bar graph indicates the frequency distribution of the first digit of the $\tau$ lepton branching fraction in terms of the fraction of total occurrences for 128 data points, irrespective of any errors present in the measurements. The values expected by Benford's law are graphed alongside, with the errorbars indicating the binomial root mean square error.}
  \label{fig:boat3}
\end{figure}

\begin{table}
\begin{center}
\begin{tabular}[t]{|l ||c|| r|}
\hline\hline
 Digit & Count & Expected \\ [0.5ex]
 \hline
 1 & 44 & 38.53 $\pm$ 5.19\\
 2 & 18 & 22.53 $\pm$ 4.30\\
 3 & 15 & 15.99 $\pm$ 3.74 \\
 4 & 14 & 12.40 $\pm$ 3.34\\
 5 & 6  & 10.13 $\pm$ 3.05\\
 6 & 6  & 8.56 $\pm$ 2.82\\
 7 & 6  & 7.42 $\pm$ 2.64\\
 8 & 6  & 6.54 $\pm$ 2.49\\
 9 & 13 & 5.86 $\pm$ 2.36\\
\hline\hline
\end{tabular}
\end{center}
\caption{128 data points are considered, irrespective of the errors present in the measurement, and the distribution of the first significant digits is determined. Similar to Table~\ref{table:1}, we observe the difference of the obtained distribution in Column 2 to the expected distribution in Column 3. Column 3 includes the expected binomial root mean square error expected for the distribution. This distribution still appears to differ from the calculated Benford's distribution, and it does not resemble the distribution in Table 1 either.}
\label{table:2}
\end{table}

It cannot be ignored that a large part of the available data on the values of the $\tau$ lepton branching fractions have high error bars, which cause a degree of uncertainty in the first significant digit of these data points. When initially compiling measurements from the data available for our earlier calculations, as discussed in Sect.~\ref{ss:data}, measurements with large errors were disregarded, in an attempt to be accurate, limiting the number of values that can be used to evaluate the similarity of the distribution to Benford's distribution.

In order to avoid bias in our results we now consider all the data points as given, irrespective  of the error, provided that we are aware of an exact calculated value of the branching fraction. Upper limiting values of branching fractions are still disregarded. Previously, Shao and Ma \cite{Shao} proceeded similarly, classifying their data into three specific categories to counter the ambiguity that their data presented them. They found similar results when they considered the questionable data as when they avoided it. Other papers~\cite{buck,ni} do not provide information on how they handled the data points with error bars, though Buck et al~\cite{buck} did refrain from testing the second digit law on experimental data points, where there was uncertainty in the value of the second significant digit.  

In a similar spirit, for completeness, we evaluate and graph the distribution of significant digits once again, this time without omitting the aforementioned data points, thus now considering 128 measurements. The resulting values are depicted in Table~\ref{table:2} and Fig.~\ref{fig:boat3} graphically represents the same.

The obtained distribution seems to deviate from Benford's law as the earlier one did. To quantify the deviation, we recalculate the Pearson $\chi ^{2}$ statistic for the newly obtained distribution. \textcolor{black}{Here, as we can see from Table~\ref{table:2}, the expected number of events in all bins is greater than 5. So for this dataset, we can use the Pearson $\chi^2$ to quantify the level of agreement.} The calculated value of $\chi ^{2}$ is equal to 13.4 for 8 degrees of freedom, giving a $p$-value of 0.097, corresponding to 1.3$\sigma$ discrepancy.

While the $p$-value is higher than that of the  previous distribution, we note that the $\chi ^{2}$ values obtained in Refs.~\cite{buck,ni,Shao}  are markedly smaller than the value calculated here, as indicated in Table~\ref{table:3}. While there does exist variance in the value of $\chi^2$ amongst the categories in these works, they do not discuss the disparity between the different datasets, instead concluding that the values of $\chi^2$ are close enough to the value of degrees of freedom that it can be assumed that Benford's law is applicable to all considered categories of the data.

\begin{table}
\begin{center}
\begin{tabular}[t]{|l ||c||c|| r|}
\hline\hline
Paper & $\chi^2$ & $\chi^2$/dof & $p$ \\ [0.5ex]
 \hline
 Buck & 5.4 & 0.60 & 0.71 \\
 Buck & 11.73 & 1.30 & 0.16\\
 Ni & 11.01 & 1.22 & 0.20\\
 Ni & 14.95 & 1.66 & 0.08\\
 Shao & 6.82  & 0.85 & 0.58\\
 Shao & 2.57  & 0.32 & 0.96\\
This work (Pearson $\chi^2$) & 21.00 & 2.62 & 0.007 \\
\textcolor{black}{This work  (Binomial  $\chi^2$)} & \textcolor{red}{16.9} & \textcolor{black}{1.87} & \textcolor{black}{0.05} \\
\hline\hline
\end{tabular}
\end{center}
\caption{Results of all previous investigations of Benford's law in particle physics along with our work \textcolor{black}{(using both Binomial and Pearson $\chi^2$)}. As each of the papers have multiple categories for which they calculated the value of $\chi^2$, we show the highest and lowest calculated value to provide comparison with the results of this work. Column 1 indicates the author and reference; Column 2 indicates the $\chi^2$ value; Column 3 is the ratio of $\chi^2$ to total degrees of  freedom; the last column is the $p$-value. Therefore, among all the particle physics datasets investigated,
only the $\tau$ branching ratios show some marginal disagreement with Benford's law.}
\label{table:3}
\end{table}


Taking into account that the previously analyzed values are more accurately known, greater dependence can be placed in the initially considered subset of the data, as opposed to this second, larger subset. It is clear that the presence of large errors affects the accuracy of the result determined, when regarding the full range of data points available for evaluation. 

\subsection{Testing the Claim}
\label{sec:basechange}

Amongst the extensive mathematical analyses performed on the significant digit phenomenon, Hill's theorems of scale invariance and base invariance have come to be accepted as comprehensive tests of Benford's Law. That is, if a dataset obeys the law, it can be shown that it's properties are invariant when the data is converted to another base, or scaled by a positive factor, as stated by Hill~\cite{Hillb}. We find that the distribution obtained for $\tau$ branching fractions varies significantly under both these tests, providing further evidence as to their disagreement with the law. We now discuss these tests.

\subsubsection{Scale Invariance}
Hill defines scale invariance as the case where a probability measure remains invariant when the data-points are multiplied by a constant, positive factor, thus rescaling them to new values. Regularly, we see this manifest as the invariance of Benford's Law under a change of units. Here, the branching fractions are dimensionless quantities and devoid of units. Thus to test scale invariance, we simply pick a set of positive, numerical factors.

We consider the original 69 values of the branching fractions for which the significant digit can be stated confidently, and proceed to scale them up by a collection of positive and rational factors. For this purpose we choose our factors to be integers from two up to nine and some non-integral factors picked randomly. For each of  these factors we re-evaluate the Pearson $\chi^{2}$ statistic, and observe the differences that we obtain in the value of $\chi ^{2}$/dof and the $p$-value. \textcolor{black}{Note that the aim of this exercise  is a proof of principles test to check the robustness of our results  with respect to a different scale. For this purpose we use the same metric (viz. Pearson $\chi^2$) for all values of the scale factors.}

The results are tabulated in Table~\ref{table:4}. We observe the large variations in the calculated statistic, confirming that the $\tau$ branching fractions do not conform to Benford's Law.

\begin{table}
\begin{center}
\begin{tabular}[t]{|l ||c||c|| r|}
\hline\hline
Factor &  $\chi^{2}$  & $\chi^{2}$/dof & $p$ \\ [0.5ex]
\hline
1			&  21.00 	&	2.62		& 	0.007	 \\
2			&  5.422 	&	0.67		& 	0.711	 \\
3			&  10.16 	&	1.27		& 	0.253	 \\
4			&  9.688 	&	1.21		& 	0.287	 \\
5			&  14.39 	&	1.79		& 	0.072	 \\
6			&  20.79 	&	2.59		& 	0.007	 \\
7			&  16.09 	&	2.01		& 	0.041	 \\
8			&  13.40 	&	1.67		& 	0.098	 \\
9			&  19.92 	&	2.49		& 	0.010	 \\
1.5		&  6.365 	&	0.79		& 	0.606	 \\
3.2		&  13.86 	&	1.73		& 	0.085	 \\
7.6		&  14.31 	&	1.78		& 	0.073	 \\
\hline\hline
\end{tabular}
\end{center}
\caption{Evaluated values of the Pearson $\chi^2$ statistic for multiple values of a scale factor. The original 69 points are each multiplied by the scale factor and the analysis is performed once again. Column 1 indicates the value of the scale factor, Column 2 the value of Pearson $\chi^2$, Column 3 is the value of $\chi^2$/dof, where in this case there are 8 degrees of freedom and Column 4 indicates the calculated $p$-value. We note the variance in the obtained results, indicating a dependence on scale.}
\label{table:4}
\end{table}

\subsubsection{Base Invariance}

\begin{figure}
  \includegraphics[width = \linewidth,scale=0.5]{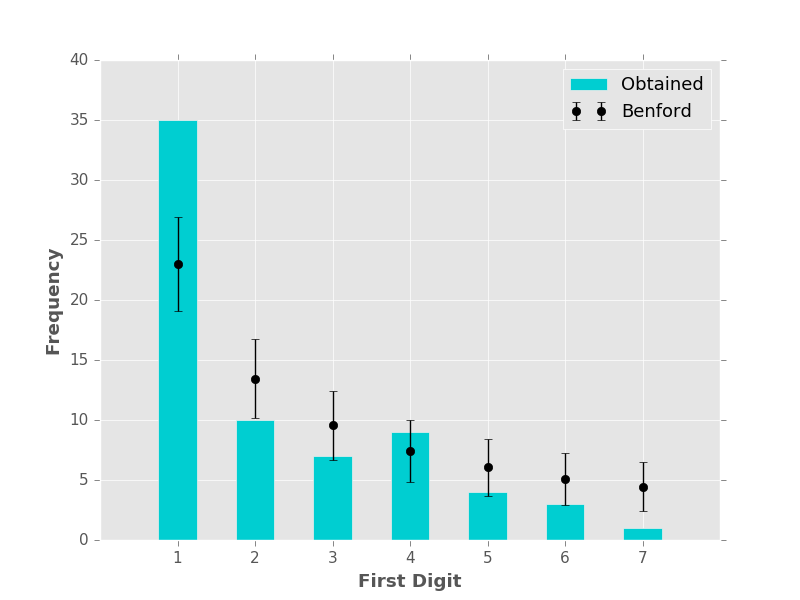}
  \caption{The bar graph indicates the frequency distribution of the first digit of the $\tau$ lepton branching fraction in terms of the fraction of total occurrences for 69 data points when represented in base 8. The values expected by Benford's law are graphed alongside, with the error bars indicating the binomial root mean square error.}
  \label{fig:boat4}
\end{figure}

Hill defines base invariance as the case where a probability measure remains unchanged when the data-points are converted from their original form such that they are represented in a different base. This property may be utilized as a strong test of Benford's Law. 

Again we consider the same 69 values of the branching fractions, and this time we convert each of the data-points so that they are rewritten in a new numerical base. We re-evaluate the Pearson $\chi ^{2}$ statistic for bases 3 to 9, and base 16. We observe the differences that we obtain in the value of $\chi^{2}$/dof and the $p$-value. For the Pearson $\chi^{2}$, degrees of freedom are given as $b-2$ where $b$ is the base of representation. 

\begin{figure}
  \includegraphics[width = \linewidth,scale=0.5]{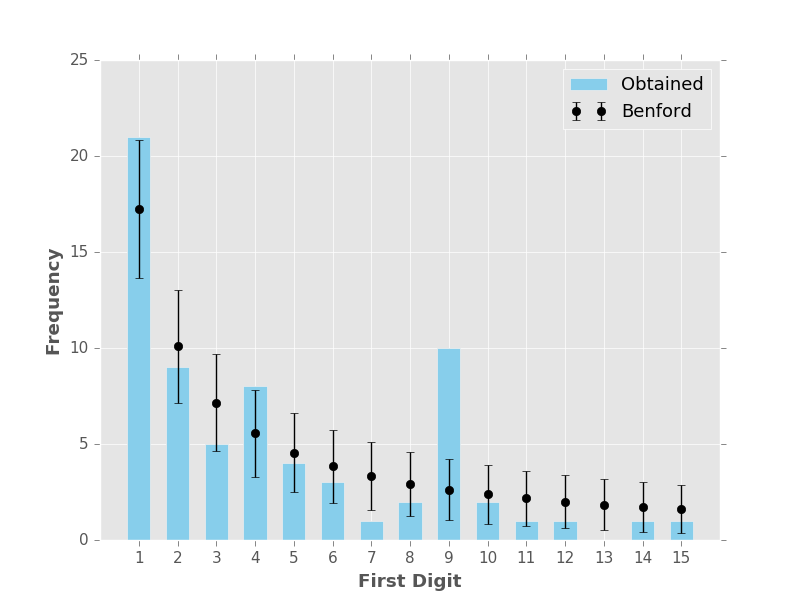}
  \caption{The bar graph indicates the frequency distribution of the first digit of the $\tau$ lepton branching fraction in terms of the fraction of total occurrences for 69 data points when represented in terms of base 16. The values expected by Benford's law are graphed alongside, with the error bars indicating the binomial root mean square error.}
  \label{fig:boat5}
\end{figure}

The results are tabulated in Table ~\ref{table:5}. Additionally, Figures ~\ref{fig:boat4} and ~\ref{fig:boat5} depict the obtained distributions for bases 8 and 16. As with the changes in scale, changes in base also cause large variations in the statistic, leading us to conclude that $\tau$ branching fractions indeed do not satisfy Benford's Law.

\begin{table}
\begin{center}
\begin{tabular}[t]{|l ||c||c|| r|}
\hline\hline
$b$ &  $\chi^{2}$  & $\chi^{2}$/dof & $p$ \\ [0.5ex]
\hline
3		&	5.576   &  5.57   &   0.018\\
4		&	3.692   &  1.84   &   0.157\\
5		&	6.442 	&  2.14		&	  0.091\\
6		&	9.283 	&  2.32		&	  0.054\\
7		&	6.454 	&  1.29		&	  0.264\\
8		&	12.39 	&  2.06		&	  0.053\\
9		&	15.43 	&  2.20		&	  0.030\\
10	&	21.00 	&  2.62		&	  0.007\\
16	&	29.14 	&  2.08		&	  0.010\\
\hline\hline
\end{tabular}
\end{center}
\caption{Evaluated values of the Pearson $\chi^2$ statistic for multiple numerical bases. The original 69 points are each changed such that they are represented in a different base and the analysis is performed once again. Column 1 indicates the value of the base $b$, Column 2 shows the value of Pearson $\chi^2$, Column 3 is the value of $\chi^2$/dof, where there are $b-2$ degrees of freedom in each case and Column 4 indicates the calculated $p$-value. We note the variance in the obtained results, indicating a dependence on base.}
\label{table:5}
\end{table}

\section{Discussions}
\label{sec:corr}

\textcolor{black}{Statistically, every calculation presented so far presents compelling evidence to suggest that the $\tau$ branching fractions show a marginal disagreement with the Benford's distribution. 
This is in slight discord with results from almost all previous data sets analyzed for compatibility with Benford’s law. 
Although, this difference should not be of concern, we now discuss one possible reason for not getting pristine agreement.}

\begin{figure}
  \includegraphics[width = \linewidth,scale=0.5]{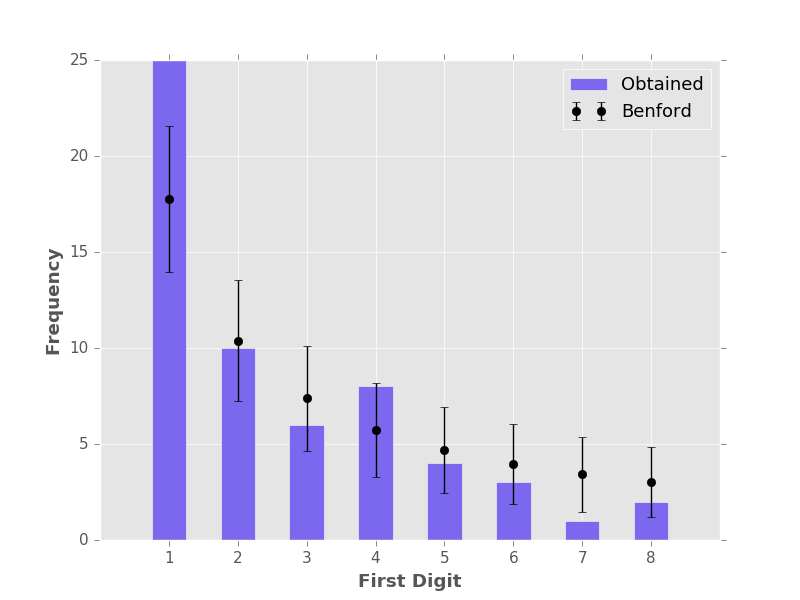}
  \caption{The bar graph indicates the frequency distribution of the first digit of the $\tau$ lepton branching fraction in terms of the fraction of total occurrences for 59 data points with low systematic errors and excluding data points, which lead with the digit 9. The values expected by Benford's law are graphed alongside, with the errorbars indicating the binomial root mean square error.}
  \label{fig:boat6}
\end{figure}

From Figs. 1-2, we can see by eye  that  the resulting distributions do tend to follow a somewhat decreasing trend as expected from Benford's law as we increase the digit, with the number of occurrences of the leading digit `1' consistently being maximum. However the number `9' consistently appears to overshoot (4$\sigma$ in Fig.~\ref{fig:boat2} and 3$\sigma$ in Fig.~\ref{fig:boat3}) its expected number of occurrences. Noting the abnormal spike in the number of occurrences of the digit `9' in our initial analysis, as shown in Fig~\ref{fig:boat2}, we revisit the original 69 points, but now omit  all of the occurrences with the leading digit `9'. The obtained distribution is depicted in Fig~\ref{fig:boat6}. As we can see from this figure, all the data points agree within $1\sigma$ of the expected values, implying perfect agreement with Benford's law if we omit the digit `9'.

However, one caveat is that we cannot ignore an entire data point, the leading digit `9' in this scenario, without a  credible, cogent reason.
Thus, our original analysis still stands as the valid obtained result. Nevertheless, this observed peculiarity of the distribution deserves some attention. We now investigate this point.

Among the different decay modes of the tau lepton, certain modes can be linked to similar underlying physical mechanisms, causing comparable values in the obtained branching fraction. Thus, there exists a certain amount of correlation among the various values of branching fraction, causing disparities in the occurrence of each leading digit.  \textcolor{black}{In this case,  among the ten occurrences of the digit `9', eight are correlated in three  bunches of 
nearly identical processes. These correspond to  $h^{-} +  2\pi^0 + \nu_t$, which occurs thrice; $h^-h^-h^+ \nu_{\tau}$,  which  also shows up three times; and $2\pi^- + \nu_t$, which is listed twice. However, in the PDG these have been reported separately as eight distinct data points, as there are additional decay products which differ for the correlated modes. If we count each group of correlated modes only once, we shall then get only five occurrences of the digit `9', instead of ten. This is consistent within $1\sigma$ of the expected number of occurrences of the digit `9' (cf. Table~\ref{table:1}) in accord with Benford's law.} 

However, there is no well defined procedure  or any literature to  disentangle the correlated datasets from the uncorrelated ones, in order to check for consistency with Benford's law. We are not aware of any previous work on Benford's law where correlated datasets have been grouped onto one or also what criterion has been used to decide if a dataset is correlated or not. \textcolor{black}{However, for this particle physics dataset, we have demonstrated that if the decay modes have the same end products, and if they are counted only once for each correlated mode, we obtain good agreement with Benford's law. Another minor discrepancy is the digit '7', which shows a $1.5\sigma$ deficit compared to expectation. However, this is not a problem given that there are  nine
bins.}




\section{Conclusions}
\label{sec:conclusions}
In this work, the first significant digit of the  $\tau$ lepton branching fractions have been tested for adherence to, and have been found to be in disagreement with Benford's law. For this purpose, the  $\chi^{2}$ statistic valid for binomial data~\cite{Cousins} was evaluated to quantify the disagreement and the ratio of $\chi^{2}$ to the total degrees of freedom was found to be 1.87, indicating  a 1.6$\sigma$  with Benford's law without including data points with large values of systematic error, and 1.68, when all available measurements were compiled regardless of error, indicating an error of 1.3$\sigma$. Furthermore, the data points without systematic errors were further tested for scale invariance and base invariance. Large fluctuations were obtained in the calculated statistical measures for changes in base and scale, providing additional evidence for the consistency of this slight disagreement with Benford's law.

As the first of the quantities in particle physics found to be in \textcolor{black}{marginal disagreement} with the law, the phenomenon suggests that the similarities in certain decay processes cause correlation amongst the data points that cause a deviation from the law. We note that there may also be some natural aspect hitherto undiscovered, that may account for the observed discrepancies. Among all the observed digits, the digit 9 shows the maximum discrepancy of about a 4$\sigma$ excess. We then examined the possible reason for this excess. 
\textcolor{black}{We then noticed that out of ten occurrences
of the digit `9', eight occur in three distinct of groups of correlated modes, where all correlated modes have two or more common decay products. If we count all correlated decay modes only once, then the number of times the digit `9' appears in the decimal digit is equal to five, which is consistent with Benford's law.}

\textcolor{black}{Therefore, this explains  the reasons for our marginal  disagreement with Benford's law, when all modes are included. However, even outside of the domain of  particle Physics,  no one has done a detailed study of  how well  Benford's law works for  correlated data, or what criterion needs to be used to check  for correlated data, and what corrections need to be made.}

Furthermore, most of the explanations of Benford's law provided up until this point fail to provide reasonable clarity as to which elementary particle datasets may obey the law, which may not. We look forward to examining these queries in the future.

\begin{acknowledgements}
We would like to thank the anonymous referees for detailed critical feedback on our manuscript. 
\end{acknowledgements}

\bibliography{tau_branch}   
\end{document}